\documentclass[conference]{IEEEtran}
\ifCLASSINFOpdf
  \usepackage[pdftex]{graphicx}
  \graphicspath{//}
\else
\fi

\usepackage{cite}
\usepackage{amsthm,amssymb}
\usepackage{svg}
\usepackage{subfig}
\usepackage{graphicx}
\usepackage{xcolor}

\usepackage{tikz}
\begin{document}
\title{Streaming and Batch Algorithms for Truss Decomposition}

\author{\IEEEauthorblockN{Venkata Rohit Jakkula}
\IEEEauthorblockA{jakku004@umn.edu\\
University of Minnesota - Twin Cities\\
Minneapolis, Minnesota\\}
\and
\IEEEauthorblockN{George Karypis}
\IEEEauthorblockA{karypis@umn.edu\\
University of Minnesota - Twin Cities\\
Minneapolis, Minnesota\\}}


%Telephone: (800) 555--1212, Fax: (888) 555--1212}


\newcommand\copyrighttext{%
  \footnotesize \textcopyright 2019 IEEE. Personal use of this material is permitted.
  Permission from IEEE must be obtained for all other uses, in any current or future
  media, including reprinting/republishing this material for advertising or promotional
  purposes, creating new collective works, for resale or redistribution to servers or
  lists, or reuse of any copyrighted component of this work in other works.}
\newcommand\copyrightnotice{%
\begin{tikzpicture}[remember picture,overlay]
\node[anchor=south,yshift=10pt] at (current page.south) {\fbox{\parbox{\dimexpr\textwidth-\fboxsep-\fboxrule\relax}{\copyrighttext}}};
\end{tikzpicture}%
}

\maketitle
\copyrightnotice

\begin{abstract}
%\boldmath
    % Detecting well connected subgraphs is an important problem, and \textit{truss decomposition} is one of the several methods used for this purpose. 
    % As the graph structure changes over time, keeping track of the changes in truss decomposition provides additional insights that domain experts can leverage. 
    % In this work, we make three significant contributions. 
    % First, we propose incremental algorithms for truss decomposition for streaming graph data. 
    % We present incremental algorithms that can update the truss decomposition after every edge insertion $-$ the algorithm works with minor modifications for edge removals as well.
    Truss decomposition is a method used to analyze large sparse graphs in order to identify successively better connected subgraphs.
	Since in many domains the underlying graph changes over time, its associated truss decomposition needs to be updated as well.
	This work focuses on the problem of incrementally updating an existing truss decomposition and makes the following three
significant contributions.
    First, it presents a theory that identifies how the truss decomposition can change as new edges get added.
    Second, it develops an efficient incremental algorithm that incorporates various optimizations to update the truss decomposition after every edge addition. These optimizations are designed to reduce the number of edges that are explored by the algorithm. 
    Third, it extends this algorithm to batch updates (i.e., where the truss
decomposition needs to be updated after a set of edges are added), which reduces the overall computations that need to be performed.
	We evaluated the performance of our algorithms on real-world datasets. Our incremental algorithm achieves over 250000$\times$ average speedup for inserting an edge in a graph with 10 million edges relative to the non-incremental algorithm. Further, our experiments on batch updates show that our batch algorithm consistently performs better than the incremental algorithm.
    % To reduce the computational complexity associated with updating the truss decomposition, our approach first explores a small set of edges whose truss numbers can change and perform the required computations only to those edges. 
    % To this end, we state theorems using which we develop an incremental algorithm similar to the one presented in Huang et al. We then build upon this to present another incremental algorithm that explores an even smaller set of edges.
    % We evaluate our algorithms on various real-world datasets. Our implementation achieves over 250000$\times$ average speedup for inserting an edge in a graph with 10 million edges relative to the non-incremental algorithm. Such speedups allow for updating the truss decomposition after every edge insertion relatively quickly, thereby making it feasible to query and analyze the graph structure at any point of time.
    % Secondly, we show that certain parts of the incremental algorithms can be executed independent of each other, thereby exposing parallelism.
    % Finally, we extend the theory developed for the incremental algorithms to propose the first batch algorithm for truss decomposition. In general, using a batch algorithm proves beneficial to using an incremental algorithm, when we are interested in finding the truss decomposition only after a batch of edge updates, as opposed to updating the truss decomposition with every edge update.
%    \textcolor{red}{Add details related to experiments.}
\end{abstract}



% \fi
\IEEEpeerreviewmaketitle


%%%%%%%%%%%%%%%%%%%%%%%%%%%%%%%%%%%%%%%%%%%%%%%%%%%%%%%%%%%%%%%%%%%%%%%%%%%%%%%%%%%%%%%%%%%%%%%%
\section{Introduction}
%%% Background
%%% Why is the notion of cohesive subgraph important? What areas is it used in, and why?
Graphs are used to represent relationships between entities, where the vertices represent the entities and the edges represent their relationships. 
For example, a social network can be represented as a graph, where the vertices represent the people, and the presence of an edge between two people denotes the existence of a relationship between them. 
Some examples of the domains in which graphs are used are telecommunications and biological systems such as in study of proteins.
% Graphs are also used in telecommunications and in biological systems such as in the study of proteins, among several other domains.
For any organization with a reasonable amount of graph data, it is often beneficial to capture the graph structure and discover important areas in the graph.
For example, finding strongly-knit communities in a social network helps in targeted advertising~\cite{fortunato2010community} whereas finding cliques in protein structure is essential for comparative modeling\cite{samudrala1998graph}.

%%% Current Knowledge
%%% What are the different types of cohesive subgraphs in the literature. Briefly explain k-truss and truss decomposition.
Several cohesive subgraphs have been proposed that capture important areas in the graph. 
The $k$-truss\cite{cohen2008trusses} is one such cohesive subgraph.
A $k$-truss of a graph $G$ is an induced subgraph of $G$ such that each edge in the subgraph is part of at least $k-2$ triangles.  Conceptually, every relationship in a $k$-truss is reinforced by the presence of at least $(k-2)$ mutual relationships in that $k$-truss. 
This makes it suitable for several applications in network science including  community detection\cite{saito2008extracting},\cite{verma2013network},\cite{huang2015approximate}, visualization\cite{alvarez2006large}, etc.
\textit{Truss decomposition} is the task of determining the maximum value of $k$ for each edge in the graph, such that the edge is part of some $k$-truss. This provides an efficient way to discover all $k$-trusses in a graph, for any value of $k$.

%%% Limitations of Existing Approaches
Most real-world graphs change as new nodes and edges are added and existing nodes and edges are removed. 
As the graph updates with time, an important question to answer in network science is how the structure of the cohesive subgraphs (like the $k$-truss) change. 
Answering this question helps in detecting significant changes in the community structure in a social network as new relationships are formed and severed. In some cases, we might be interested in how the community structure looked like at a certain point in the past. 
When the cohesive subgraphs are $k$-trusses, such questions can be answered by performing \textit{truss decomposition} after each update. 
While there are several serial and parallel algorithms for truss decomposition\cite{wang2012truss},\cite{chen2014distributed},\cite{smith2017truss},\cite{kabir2017parallel}, these algorithms explore the entire graph. 
As a result, performing truss decomposition after every update can become computationally expensive. 
However, since most updates would affect the structure of only those communities in the proximity of the edge being inserted/removed, the changes in the truss decomposition will tend to be localized around the area of the graph in which the change occured.
Huang et al.\cite{huang2014querying} builds upon this intuition to present an incremental algorithm for truss decomposition; however, this algorithm checks more edges than necessary to see if they are affected due to an update.
% Batch Updates
In other cases, we might be interested in finding the truss decomposition after a batch of edge updates. While an incremental algorithm could be used to perform batch updates, there is a possibility of redundant computations being performed over a batch of edges. To the best of our knowledge there is no batch algorithm for truss decomposition that handles this problem.

%%% Paper's smart solution and its contributions
% Connect the limitations mentioned in the previous section
In this paper, we make the following contributions. First, we build on the work of Huang et al.\cite{huang2014querying} to develop a theory that provides an upper bound to the subset of the edges that need to be explored, such that the change in truss decomposition due to an update is guaranteed to contain within this subset.
% Incremental algorithms
Using this, we develop an efficient incremental algorithm that explores a smaller set of edges as compared to the algorithm developed by Huang et al.~\cite{huang2014querying}.
% Parallel components
Furthermore, we show that our algorithm exhibits a high degree of concurrency, which can be exploited by a parallel formulation.
% Batch updates
Finally, we extend the theory used to develop the incremental algorithm to efficiently perform batch updates. Using the batch algorithm, we can update the truss decomposition after a batch of edge updates faster than updating the truss decomposition after every edge update. 
% This is useful in cases where we are not interested in finding the truss decomposition after every edge update, but only after a batch of edge updates. 
Note that our work considers the problem of updating the truss decomposition only when the stream consists of edge insertions $-$ the theory can be extended to edge removals as well.

% Results
We evaluated the performance of our algorithms on a sparse and a dense real-world dataset. We test our algorithms for scalability by simulating a streaming scenario at different sizes of the underlying graph. 
The experiments we performed show that the incremental algorithm provides upto 250000$\times$ speedup when compared to using the non-incremental algorithm for performing edge insertion. 
Moreover, our incremental algorithm performs better than the algorithm presented by Huang et al. in most cases. 
Finally, our experiments show that the batch algorithm consistently performs better than the incremental algorithm for batch updates, running upto 17.5 times faster in some cases.

%%% Roadmap
% The rest of the paper is organized as follows. In \textsc{Section II}, we provide the necessary background and notation required to understand the sections that follow.
%In \textsc{Section III}, we provide a brief overview of some related research done in this field. 
%In \textsc{Section IV}, we state the theorems that the incremental algorithms use, and in \textsc{Section V} we explain the incremental algorithms themselves in detail.
%In \textsc{Section VI}, we present the batch algorithm.
%We detail the experiments we performed to evaluate our algorithms in \textsc{Section VII}, and follow it up with the discussion of results in \textsc{Section VIII}.
%Finally, in \textsc{Section IX}, we conclude the paper and propose future directions to our work, that can address some of its limitations. \\

%%%%%%%%%%%%%%%%%%%%%%%%%%%%%%%%%%%%%%%%%%%%%%%%%%%%%%%%%%%%%%%%%%%%%%%%%%%%%%%%%%%%%%%%%%%%%%%%
\section{Background and Notation}
% Basic Definitions
Let $G=(V,E)$ be an undirected and unweighted graph with no self-loops, where $V$ and $E$ are the vertex and edge set respectively. 
A set of vertices $\{u,v,w\} \subseteq V$ form a triangle if and only if $\{ (u,v),(v,w),(u,w) \} \subseteq E$. 
Let $Adj(v)$ denote the vertices adjacent to $v$ in G. We define the support of an edge $e=(u,v) \in E$ in the graph G as $sup_G(e) = |Adj(u) \cap Adj(v)|$. Equivalently, $sup_G(e)$ is the number of triangles that include the edge $e=(u,v)$, since $w \in Adj(u) \cap Adj(v)$ if and only if $\{(u,v), (v,w), (u,w)\} \subseteq E$. 
Moreover, we say a triangle $\Delta = \{u,v,w\}$ is supported by an edge $e=(a,b)$ if and only if $a \in \Delta$ and $b \in \Delta$.

%\begin{figure}
%  \centering
%  \includesvg[scale=0.65]{images/ktrussExample1.svg}
%  \caption{Example graph to demonstrate that an edge could belong to multiple trusses. The edge $e$ forms a $3$-truss with the edges in blue, while it forms a $4$-truss with the edges in red.}
%  \label{fig1}
%\end{figure}

% k-truss definition
We now define the notion of a $k$-truss. A $k$-truss of the graph $G$ is an induced one-component subgraph $G'$ of $G$ such that each edge in $G'$ supports at least $k-2$ triangles. In other words, for every edge $e$ in the $k$-truss $G'$, $sup_{G'}(e) \geq (k-2)$. 
It follows from the definition of a $k$-truss that if an edge is part of a $k$-truss, then it is also a part of a $k'$-truss, for all $2 \leq k'<k$. 
Moreover, each edge could possibly be a part of multiple trusses with different $k$ values. For each such value of $k$, let $H_{e,k}$ denote the $k$-truss that $e$ is a part of. 
% Example...
% For the graph in Figure \ref{fig1}, the edge $e$ forms $H_{e,4}$ (a 4-truss) with the edges in the \textit{red}, while it forms $H_{e,3}$ (a $3$-truss) with the edges in \textit{blue}.

% Definition of Truss Number of an edge
The \textit{maximal} value of $k$ for which an edge $e$ is part of a $k$-truss is called the \textit{truss number} of the edge $e$ and is denoted by $K(e)$. We denote the corresponding \textit{maximal} $k$-truss that contains the edge $e$ by $H_e$. Then we have $H_e = H_{e,K(e)}$. 
% For example, in Figure \ref{fig1}, $K(e)=4$, and $H_e=H_{e,4}$ is the induced subgraph consisting of the \textit{red} edges and the edge $e$. 

Further, we note that every edge in a $k$-truss has $K(.) \geq k$, where $K(.)$ is used to denote the \textit{truss-number} of any arbitrary edge in the $k$-truss. In general, we will use $K(.)$ to denote the truss-number of any arbitrary edge, depending on the context. 
We will use \textit{ktmax} to denote the maximum $K(.)$ value across all edges in the graph.

% Min-truss definition
Given a triangle $\Delta = \{u,v,w\}$ and an edge $e$ of the triangle (i.e., an edge with its vertices in the triangle $\Delta$), we now define the \textit{min-truss number} of the triangle $\Delta$ with respect to the edge $e$. 
Without loss of generality, let us assume $e=(u,v)$. Then, the \textit{min-truss number} of the triangle $\Delta$ with respect to the edge $e$ is defined as $\Phi(\Delta, e) = \min(K((u,w)), K((v,w)))$. 
We will use the notion of \textit{min-truss number} when we provide the implementation details of the incremental algorithm in \textsc{Section V}.

% Usage of "+" super-script
While developing the incremental algorithm, we make observations on the structural changes to the graph when an edge is inserted. 
In general, we will use the superscript $+e$ when we refer to an instance of the graph after the insertion of the edge $e$. 
In particular, while $H_{e'}$ denotes the \textit{maximal} $k$-truss that contains the edge $e'$ before $e$ is inserted, we use $H_{e'}^{+e}$ to denote the \textit{maximal} $k$-truss that contains $e'$ after the edge $e$ is inserted. 
Similarly, for a given $k$, we use $H_{e',k}^{+e}$ to denote the $k$-truss that contains the edge $e'$ after the edge $e$ is inserted, while $H_{e',k}$ refers to the $k$-truss that contains $e'$ before $e$ is inserted.

% \begin{figure}
%  \centering
%  \includesvg[scale=0.65]{images/ktrussBeforeAfter.svg}
%  \caption{Example graph before and after the insertion of edge $e$: (a) Before insertion of $e$, the edges in green form $H_p$ (a $3$-truss), the \textit{maximal} $k$-truss containing the edge $p$, (b) After insertion of $e$, $H_p$ this gets updated to $H_p^{+e}$ (a $4$-truss), which is the union of $e$ and the edges in \textit{red}.}
%  \label{fig2}
%\end{figure}

% Example
% Consider for example, the graph shown in Figure \ref{fig2} before and after the insertion of the edge $e$. Before the insertion of $e$, the edges in green form $H_{p}$, the \textit{maximal} $k$-truss that contains the edge $p$. 
% After the edge $e$ is inserted, $e$ forms a 4-truss with the \textit{red} edges, while it forms a 3-truss with the \textit{blue} edges. 
% Also, the \textit{maximal} $k$-truss for edge $p$ is updated from $H_{p}$ to $H_{p}^{+e}$, which is now the induced subgraph consisting of the edges in \textit{red} and the inserted edge $e$. 
% Note that $H_{p}^{+e} \neq H_{p}$ \textit{if and only if} $e \in H_{p}^{+e}$, since the only change in the graph is the addition of the edge $e$. \\

%%%%%%%%%%%%%%%%%%%%%%%%%%%%%%%%%%%%%%%%%%%%%%%%%%%%%%%%%%%%%%%%%%%%%%%%%%%%%%%%%%%%%%%%%%%%%%%%
\section{Previous Work}
% General intro to cohesive subgraphs
In this section, we provide a brief literature review of the existing algorithms for finding cohesive subgraphs in a graph. 
The basic form of cohesive subgraphs is the \textit{clique}, which is a subset of vertices that forms a complete subgraph. Bron et al.\cite{bron1973algorithm} provides an algorithm to compute all cliques in an undirected graph. 
The definition of clique is often too rigid and other cohesive subgraphs like $n$-clique\cite{luce1950connectivity}, $n$-clan\cite{mokken1979cliques} and $n$-club\cite{mokken1979cliques} were proposed. However, the computation of all these subgraphs is NP-hard. 

% k-core
There exist other forms of cohesive subgraphs which can be computed in polynomial time. A $k$-core\cite{zhang2012extracting} is a maximal induced subgraph in which every vertex has degree of at least $k$. The \textit{core decomposition} discovers all $k$-cores (for all possible $k$ values) in the graph. Linear time algorithms\cite{batagelj2003m} have been developed to perform core decomposition. 

% k-truss
A $k$-truss captures more important areas of the graph as compared to a $k$-core $-$ every $k$-truss is a $k$-core, but the vice-versa is not necessarily true. The first algorithm for truss decomposition was introduced by Cohen\cite{cohen2008trusses}. Several other serial, parallel and distributed algorithms have been proposed for truss decomposition. Cohen\cite{cohen2009graph} and Chen et al.\cite{chen2014distributed} provide distributed algorithms for truss decomposition. Wang et al.\cite{wang2012truss} proposes I/O efficient algorithms to handle massive networks that do not fit in main memory. Smith et al.\cite{smith2017truss} and Kabir et al.\cite{kabir2017parallel} provide efficient parallel algorithms in shared memory and distributed memory systems, respectively.  

% Nucleus decomposition

% Streaming algorithms
While the literature has several efficient algorithms for core decomposition and truss decomposition, there has been limited work done in the area of streaming algorithms for these problems. 
Sariyuce et al.\cite{sariyuce2013streaming} propose incremental algorithms for core decomposition for streaming graph data. 
Huang et al.\cite{huang2014querying} present an algorithm for incrementally updating the truss decomposition for streaming graph data.
In the following section, we extend the theoretical findings presented by Sariyuce et al.\cite{sariyuce2013streaming} to the problem of truss decomposition and develop incremental algorithms for the same.
We then extend the theory used to develop the incremental algorithms to develop the first batch algorithm for truss decomposition. \\

%%%%%%%%%%%%%%%%%%%%%%%%%%%%%%%%%%%%%%%%%%%%%%%%%%%%%%%%%%%%%%%%%%%%%%%%%%%%%%%%%%%%%%%%%%%%%%%%
\section{Theoretical Basis for Incremental Algorithms}
Employing the non-incremental algorithm to compute the $k$-truss decomposition from scratch for each edge insertion requires exploring every edge of the graph per insertion. This is computationally wasteful if an inserted edge affects the $K(.)$ values of only a small portion of the graph. 
Thus, instead of having to explore \textit{every} edge of the graph, we wish to explore a smaller portion of the graph that is guaranteed to contain \textit{all} the edges whose $K(.)$ values increase.
The theorems below help us explore only a subset of edges whose $K(.)$ values \textit{can potentially} change due to the insertion of $e$. These theorems have been rigorously proved, although we do not present them here due to space limitations.\\
\begin{itemize}
    \item 
    \underline{\textsc{Theorem 1.}} If an edge $e$ is inserted into $G$, then the $K(.)$ value of any edge can increase by at most 1.\\

    \item
    \underline{\textsc{Theorem 2.}} If an edge $e$ is inserted into $G$, then for every other edge whose $K(.)$ value increases from $k$ to $k+1$, it forms a triangle with either $e$ or with at least one other edge whose $K(.)$ value also increases from $k$ to $k+1$.
    
    This theorem provides a recursive structure to the change in truss decomposition $-$ specifically, to the edges whose $K(.)$ values increase from $k$ to $k+1$. We use this recursive structure to further state the theorem below. \\

    \item 
    \underline{\textsc{Theorem 3.}} If an edge $e=(u,v)$ is inserted into $G$, then for every edge $e'$ whose $K(.)$ value increases from $k$ to $k+1$ there exists a path $p$ in $G$ such that 
    \begin{enumerate}
        \item $e' \in p$
        \item $\forall e'' \in p, K(e'')$ increases from $k$ to $k+1$
        \item $\exists e'' \in p$ such that $e'$ and $e''$ are part of a triangle, and $K(.) \geq k+1$ for all edges of the triangle
        \item $\exists (x,y) \in p$ such that:
        \begin{enumerate}
            \item $x \in \{u, v\}$, and
            \item WLOG assume $x=u$ above. Then $(v,y) \in E$ with $K((u,y)) \leq K((v,y))$.
        \end{enumerate}
    \end{enumerate}
	\iffalse    
    \begin{proof} \textsc{Theorem 2} defines a recursive structure to the change in $K(.)$ from which the existence of a path $p$ that satisfies the properties $1$, $2$, $3$ and $4(a)$ is evident. Fig. \ref{fig4} shows such a path $p$ and illustrates properties $1$, $2$ and $4(a)$.
    
    Since $\forall e'' \in p$, $H_{e''}^{+e}$ is a $(k+1)$-truss and includes the edge $e$, $H_{e''}^{+e} = H_{e,k+1}^{+e}$. In other words, all edges $e''$ on the path $p$ belong to the same $(k+1)$-truss: $H_{e,k+1}^{+e}$. 

    If the recursive structure ends at $\tilde{e}=(x,y)$ (see Fig. \ref{fig4}), then we know that $K(\tilde{e})$ increases due to increase in the support of $\tilde e$ by the edge $e$. Assuming $x=u$ WLOG, $(v,y) \in E$, and the triangle $\{ u,v,y \} \in H_{e,k+1}^{+e}$. Since $(v,y) \in H_{e,k+1}^{+e}$, $K((v,y)) \geq k+1 = K((u,y))$. Hence, $K((u,y)) \leq K((v,y))$ and the theorem follows.
    \end{proof}
    \fi
    \underline{Note:} If $K((u,y)) = K((v,y))$ then $K((v,y))$ will increase if and only if $K((u,y))$ increases.\\
\end{itemize}

\begin{figure}[t]
	\centering
	\includegraphics[scale=0.7]{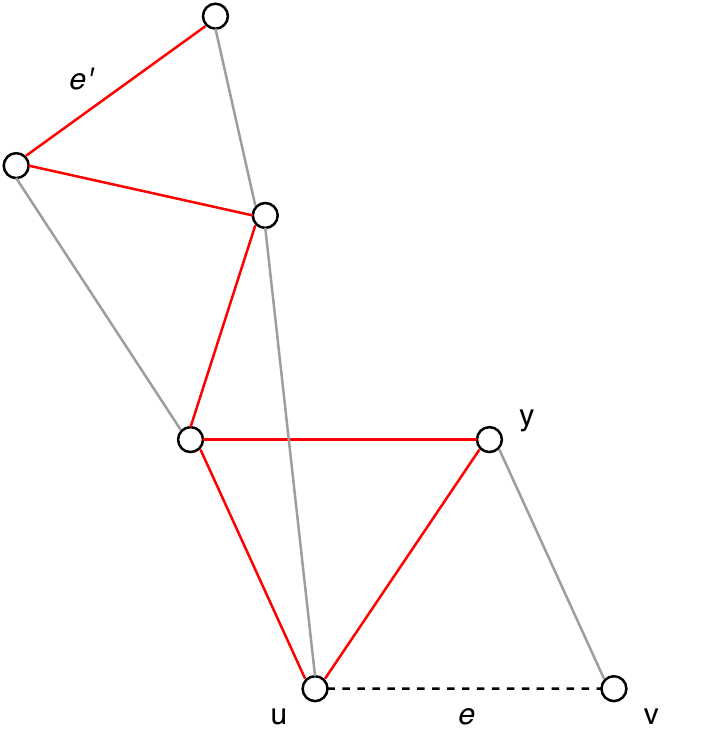}
	\caption{Portion of an example graph to demonstrate \textsc{Theorem 3}. The edges in \textit{red} denote the edges whose $K(.)$ values increase from $k$ to $k+1$, when the edge $e$ is inserted.}
	\label{fig3}
\end{figure}

Figure \ref{fig3} demonstrates what is stated in \textsc{Theorem 3}. In the example, we consider an edge $e'$ whose $K(.)$ value increases from $k$ to $k+1$ upon insertion of the edge $e$. The edges in \textit{red}, starting from $e'$, depict the recursive structure stated in \textsc{Theorem 2} and denote the edges whose $K(.)$ values increase from $k$ to $k+1$ $-$ properties 1, 2 and 3 of \textsc{Theorem 3} directly follow from \textsc{Theorem 2}. The recursive structure ends with the edge $(u,y)$, which forms a triangle with the inserted edge $e$.

%%%%%%%%%%%%%%%%%%%%%%%%%%%%%%%%%%%%%%%%%%%%%%%%%%%%%%%%%%%%%%%%%%%%%%%%%%%%%%%%%%%%%%%%%%%%%%%%
\section{Incremental Algorithms}
The above theorems help us explore a subset of edges in $E$ whose $K(.)$ values \textit{can potentially} increase due to the addition of a new edge $e=(u,v)$.
For each value of $k \in [2, ktmax]$, we explore all the paths starting from $u$ and $v$ such that each path is composed entirely of edges with $K(.)=k$, and adheres to the properties in \textsc{Theorem 3}. 

Specifically, for each triangle $\Delta$ supported by $e$, if $\Phi(\Delta, e) = k$, we pick the edge(s) of $\Delta$ with $K(.)=k$ and start exploring paths starting from these edge(s). We recursively explore paths such that $K(e')=k$ for every edge $e'$ on the path, and there exists $e'' \in p$ such that $e'$ and $e''$ form a triangle and $K(.) \geq k$ for all edges of the triangle.
The contrapositive of \textsc{Theorem 3} then guarantees that any edge outside this path will not have their $K(.)$ increased to $k+1$.
% For each value of $k \in [2, ktmax]$, \textsc{Theorem 3} suggests to look for the edges with $K(.)=k$ on a path (with properties as defined in the theorem) starting from $u$ or $v$, and guarantees that any edge outside this path will not have their $K(.)$ increased to $k+1$.

\subsection{Basic idea}
In this section, we discuss the basic idea behind the incremental algorithm, and we gloss over the finer details, which will be discussed in later sections.

Consider the set of edges, say $S$, with $K(.)=k$ that are explored using \textsc{Theorem 3}. The set of edges $S'$ whose $K(.)$ values do increase to $k+1$ upon insertion of $e$ is a subset of $S$.
Moreover, since the only change in the graph is the inserted edge $e$, for every edge $e' \in S'$, $e$ must be in the \textit{maximal} $k$-truss for $e'$. Since the \textit{maximal} $k$-truss for $e'$ is also the $(k+1)$-truss of $e$, we have $e \in H_{e'}^{+e} = H_{e,k+1}^{+e}$. Therefore, the edges in $S'$ and $e$ are part of $H_{e,k+1}^{+e}$, the $(k+1)$-truss that contains the edge $e$.
%In other words, in the subgraph $H_{e,k+1}^+$, each edge $e' \in S'\cup e$ supports at least $(k+1)-2 = k-1$ triangles. \\

Our goal is to find this set $S' \subseteq S$. For this, we need to prune the set $S$ to the subset $S'$ such that the edges in $S'$ along with $e$ and a set of other edges, say $Q$, form a subgraph, where each edge supports at least $(k+1)-2 = k-1$ triangles. Then, the subgraph satisfies the condition for it to be a $(k+1)$-truss, and we can increase the $K(.)$ value of each edge in $S'$ to $k+1$. Regarding the $K(.)$ value of the inserted edge $e$, we can only conclude that it belongs to a $(k+1)$-truss and thus $K(e)$ is at least $k+1$ $-$ determining the exact value of $K(e)$ will be discussed in later sections. Moreover, since every edge in a $(k+1)$-truss has $K(.)>k$, the set of edges $Q$ must have $K(.)>k$. 

We state the following property to summarize the above observations, which we will frequently refer to during our discussion. \\

\underline{\textsc{Property 1:}} The set of edges $S$ with $K(.)=k$ have their $K(.)$ values increase to $k+1$ \textit{if and only if} the edges in $S$ and the inserted edge $e$, together with a set $Q$ of edges with $K(.)>k$ form a subgraph where each edge supports at least $(k-1)$ triangles. \\

If the initial set of edges $S$ explored using \textsc{Theorem 3} satisfies the \textsc{Property 1}, then we are done. If not, it means that for the given set $S$, there is no set $Q$ such that \textsc{Property 1} is satisfied. To resolve this issue, we need to remove some edges from $S$ till we find the set $S' \subseteq S$ that satisfies \textsc{Property 1}.
The process of finding these edges, and the order in which we remove them to prune the set $S$ to $S'$ will be discussed in a later section.

We now have an overview of how the incremental algorithm explores a set of edges with $K(.)=k$ and prunes this set to the exact set of edges whose $K(.)$ will increase to $k+1$. Let us call this operation \textsc{AlgorithmX}($k$) for a particular value of $k$. 
The incremental algorithm needs to perform \textsc{AlgorithmX}($k$) for every $k \in [2, ktmax]$. In this section we discussed \textsc{AlgorithmX}($k$) for a given value of $k$, but did not discuss if and how executing \textsc{AlgorithmX}($k'$) for $k' \neq k$ affects the results of \textsc{AlgorithmX}($k$). 
For example, if $ktmax$ is 4, and an edge $e$ is inserted, we need to perform \textsc{AlgorithmX}($2$), \textsc{AlgorithmX}($3$) and \textsc{AlgorithmX}($4$) to find the edges whose $K(.)$ values will increase to 3, 4 and 5 respectively. However, we do not know if the order of executing these will affect the results. 
%We will show in the following section that the order of performing \textsc{AlgorithmX}($k$) for different values of $k$ does not matter, and each \textsc{AlgorithmX}($k$) can be executed independently.

\subsection{Order of executing \textsc{AlgorithmX}($k$) for different values of $k$}
Let us consider \textsc{AlgorithmX}($k$). We will show that executing \textsc{AlgorithmX}($k'$) where $k' \neq k$ does not affect \textsc{AlgorithmX}($k$). 
%Once we show this, then we know that \textsc{AlgorithmX}($k$) can be executed independent of \textsc{AlgorithmX}($k'$), and hence conclude that the order of the execution of \textsc{AlgorithmX}($k$) for different values of $k$ does not matter. We will show this case-by-case as follows.
\begin{enumerate}
    \item 
    \underline{Case 1:} \textsc{AlgorithmX}($k'$) does not affect \textsc{AlgorithmX}($k$) when $k' > k$.
    
    When \textsc{AlgorithmX}($k'$) is executed, the only edges whose $K(.)$ values increase are those with $K(.)=k'$.
    Since $k'>k$, the set of edges $Q$ that satisfy \textsc{Property 1} for the set $S'$ while executing \textsc{AlgorithmX}($k$) will not change irrespective of whether \textsc{AlgorithmX}($k'$) is executed or not. Moreover, if \textsc{Property 1} does not hold for a set $S$ when executing \textsc{AlgorithmX}($k$) before \textsc{AlgorithmX}($k'$), it will not hold even after executing \textsc{AlgorithmX}($k'$), again owing to the fact that $k'>k$.
    
    In conclusion, when executing \textsc{AlgorithmX}($k$) before \textsc{AlgorithmX}($k'$) for any $k'>k$, \textsc{Property 1} holds for a set $S$ of edges with $K(.)=k$ \textit{if and only if} \textsc{Property 1} also holds when executing \textsc{AlgorithmX}($k$) after \textsc{AlgorithmX}($k'$).
    
    \item
    \underline{Case 2:} \textsc{AlgorithmX}($k'$) does not affect \textsc{AlgorithmX}($k$) when $k' < k$.
    
    When \textsc{AlgorithmX}($k'$) is executed, the only edges whose $K(.)$ values increase are those with $K(.)=k'$. Since $k'<k$, the edges that are affected by \textsc{AlgorithmX}($k'$) will have their $K(.)$ values updated to no more than $k$. These affected edges cannot have their $K(.)$ values increase again, due to \textsc{Theorem 1}. Therefore, the edges whose $K(.)$ values increase to $k$ during \textsc{AlgorithmX}($k'$) cannot be part of $S'$ when executing \textsc{AlgorithmX}($k$).
    
    Moreover, since the set $Q$ of edges that satisfy \textsc{Property 1} for the set $S'$ while executing \textsc{AlgorithmX}($k$) has edges with $K(.)>k$, the edges affected by \textsc{AlgorithmX}($k'$) have no role to play in \textsc{AlgorithmX}($k$). 
    
    As a result, the edges affected by \textsc{AlgorithmX}($k'$) cannot be a part of $S'$ or $Q$ while executing \textsc{AlgorithmX}($k$), and we conclude that \textsc{AlgorithmX}($k'$) does not affect \textsc{AlgorithmX}($k$) when $k' < k$.
\end{enumerate}
This is an important observation we make in this paper. Since \textsc{AlgorithmX}($k$) can be executed independent of other \textsc{AlgorithmX}($k'$), where $k' \neq k$, this exposes parallelism which can be exploited. Due to limitations of time, we do not exploit this parallelism in this paper.

\subsection{Prune $S$ to $S'$ during \textsc{AlgorithmX}($k$)}
%In \textsc{Section A}, we gave an overview of how \textsc{AlgorithmX}($k$) works by pruning the set $S$ to the set $S'$ such that \textsc{Property 1} is satisfied. 
In this section, we will discuss \textit{how} to prune the set $S$ to $S'$. It follows from \textsc{Property 1} that we can increase the $K(.)$ values of all edges in a set $S$, \textit{if and only if} there exists a set $Q$ of edges with $K(.)>k$ such that the subgraph with edges $S \cup Q \cup \{e\}$ forms a $(k+1)$-truss.
In other words, we cannot increase the $K(.)$ values of the edges in a set $S$, \textit{if and only if} there doesn't exist a set $Q$ of edges with $K(.)>k$ such that every edge in $S \cup Q \cup \{e\}$ supports at least $(k+1)-2=(k-1)$ triangles where each triangle is composed of edges that either belong to $Q$ (whose edges have $K(.)>k$) or $S \cup \{e\}$. 

Since every edge in $Q$ has $K(.)>k$ and therefore belongs to at the least a $(k+1)$-truss, it is always possible to add edges to the set $Q$ such that every edge in $Q$ supports at least $(k-1)$ triangles where each triangle is composed of edges that belong to $Q$.
With this observation, we can further restate \textsc{Property 1} as follows: we cannot increase the $K(.)$ values of all edges in a set $S$, \textit{if and only if} there doesn't exist a set $Q$ of edges with $K(.)>k$ such that every edge in $S \cup \{e\}$ supports at least $(k-1)$ triangles where each triangle is composed of edges that either belong to $Q$, or $S \cup \{e\}$. 

This leads to the following idea: Given a set of edges $S$, we check if every edge $e' \in S \cup \{e\}$ supports at least $(k-1)$ triangles such that each triangle is composed of edges that either have $K(.)>k$, or belong to $S \cup \{e\}$. 
\begin{enumerate}
    \item
    If this is true, then we can let $Q$ be those edges with $K(.)>k$, and using the earlier observation, add more edges to $Q$ such that every edge in $Q$ supports at least $(k-1)$ whose edges are in $Q$ as well. 
    Then $S$ satisfies \textsc{Property 1} and we can increase the $K(.)$ values of all edges in $S$ to $k+1$.
    \item
    If this is not true, we remove the edges in $S$ which do not support at least $(k-1)$ triangles with the required property $-$ each triangle is composed of edges that either have $K(.)>k$ or belong to $S \cup \{e\}$. 
    Removal of edges in $S$ could reduce the support of other edges in $S$, leading to a cascading effect $-$ we continue removing the edges from $S$, till the required property holds for all edges in $S$.
\end{enumerate}

%\subsection{Implementation}
% \textsc{Section C} described the process of pruning the set $S$ of edges with $K(.)=k$ that were explored using \textsc{Theorem 3}, to the required set $S'$ whose edges actually have their $K(.)$ values increase due to the inserted edge $e$. 
%In this section, we will give provide the implementation details that are required for the pruning operation during \textsc{AlgorithmX}($k$).

It follows that we need to keep track of the number of triangles supported by each edge in $S \cup \{e\}$ such that each triangle is composed of edges that either have $K(.)>k$ or belong to $S \cup \{e\}$. 
When we start with the initial set $S$ (when we have not removed any edges yet), this is equivalent to counting the number of triangles supported by each edge in $S \cup \{e\}$ such that each triangle is composed of edges with $K(.) > k$ or $K(.) = k$. 
This is because for every edge in $e' \in S$, every other edge with $K(.) = k$ that forms a triangle with $e'$ such that the triangle is composed of edges with $K(.) \geq k$, is also in $S$, by construction. Therefore, for each edge $e'$ in the initial set $S \cup \{e\}$, we count the number of triangles supported by $e'$ such that for each triangle $\Delta$, $\Phi(\Delta, e') \geq k$. We will call this count as the \textit{relevant support count}.

If the \textit{relevant support count} is at least $k-1$ for all $e'$, then $S$ is the required set that satisfies \textsc{Property 1}, and we are done.
Otherwise, we pick the edges for which the \textit{relevant support count} is less than $k-1$, and remove those from the set one after the other. At this point, it is worth noting that a \textit{necessary condition} for an edge to have its $K(.)$ increase to $k+1$ is that its \textit{relevant support count} be at least $k-1$. For each edge that we remove from the set $S$, we also update (decrease by 1) the \textit{relevant support count} of the other edges in $S$ that lose support due the removal of the edge. This lets us maintain for each edge in $S \cup \{e\}$, the count of the number of triangles in $S \cup \{e\}$ such that each triangle is composed of edges that have either $K(.)>k$ or belong to $S \cup \{e\}$.

Once the set $S$ has been pruned to the set $S'$ such that the \textit{relevant support count} of every edge in $S' \cup {e}$ is at least $k-1$, then we can increase the $K(.)$ of every edge in $S'$ to $k+1$, and \textsc{AlgorithmX}($k$) is completed.

Moreover, since we know that a \textit{necessary condition} for an edge to have its $K(.)$ increase to $k+1$ is that its \textit{relevant support count} be at least $k-1$, if the \textit{relevant support count} for the inserted edge $e$ is less than $k-1$ for a particular value of $k$, $e$ cannot be a part of a $(k+1)$-truss. As a result, whenever the \textit{relevant support count} of $e$ for a value of $k$ is less than $k-1$, we do not execute \textsc{AlgorithmX}($k$) for that value of $k$.

\subsection{Determine $K(.)$ value of the inserted edge}
%\textsc{Section D} provides the implementation details required to perform \textsc{AlgorithmX}($k$) for all $k \in [2, ktmax]$.  

To complete the algorithm, we need to finally calculate the $K(.)$ of the inserted edge $e$. 
We know that the edge $e$ belongs to a $k$-truss, \textit{only if} $e$ supports at least $k-2$ triangles such that for each triangle $\Delta$, $\Phi(\Delta, e) \geq k$. 
After executing \textsc{AlgorithmX}($k$) for all valid $k \in [2, ktmax]$, every edge except $e$ that is affected by the inserted edge is updated, and $K(e)$ is simply the largest value of $k$ for which the $e$ supports at least $(k-2)$ triangles with the other two edges having $K(.) \geq k$. 
Equivalently, $K(e)$ is simply the largest value of $k$ for which the \textit{relevant support count} of $e$ is at least $k-2$.

The above discussion presents the details of an incremental algorithm for truss decomposition.
This algorithm is similar to the one presented in Huang et al.\cite{huang2014querying} and hence we will call this as the \textit{HCQTY} version (following from the initials of the authors). 
We approach the problem differently when compared to Huang et al. and provide additional insights into the incremental algorithm. 
In particular, our approach shows that certain parts of the algorithm can be executed parallely, and as we will see in coming sections, the theory we developed can be easily extended to a batch algorithm for truss decomposition. 

\subsection{Improved version of the incremental algorithm}
In this section, we build on the \textit{HCQTY} version of the algorithm and incorporate certain optimizations. We will call this algorithm as the \textit{JK-Inc} version (again following from the initials of authors).

We made a crucial observation in \textsc{Section C} regarding the \textit{relevant support count} $-$ while executing \textsc{AlgorithmX}($k$), the only edges in the initial set $S$ that are of interest to us are the edges with the \textit{relevant support count} at least $k-1$. We can pre-compute these counts for each edge of the graph $-$ we will call this count for an edge as it \textit{truss-degree}, and we will redefine the \textit{relevant support count} for \textit{JK-Inc} version of the incremental algorithm. As mentioned before, the only edges we are interested in are those with \textit{truss-degree} at least $k-1$. Therefore, we define the \textit{relevant support count} of an edge in $S$ as the number of triangles supported by the edge, such that either $\Phi(\Delta, e) > k$, or $\Phi(\Delta, e) = k$ and the edges of the triangle $\Delta$ with $K(.)=k$ have \textit{truss-degree} at least $k-1$.

This redefinition of \textit{relevant support count} and \textit{truss-degree} motivated from Sariyuce et al.\cite{sariyuce2013streaming} effectively reduces the number of edges we explore in the initial set $S$, thereby reducing the total number of computations. However, since we pre-computed the \textit{truss-degree} values, we need to recompute these before we perform the next update. This can be done efficiently, since the only edges whose \textit{truss-degree} needs to be recomputed are those that belong to triangles whose other edge(s) had their $K(.)$ value increase during the incremental algorithm.

%%%%%%%%%%%%%%%%%%%%%%%%%%%%%%%%%%%%%%%%%%%%%%%%%%%%%%%%%%%%%%%%%%%%%%%%%%%%%%%%%%%%%%%%%%

\section{Batch Algorithm}
\begin{figure}[h]
	\centering
	\includegraphics[scale=1]{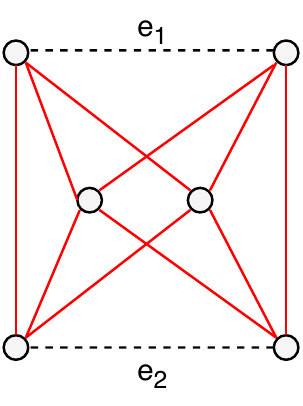}
	\caption{Example graph to demonstrate the batch algorithm}
	\label{fig4}
\end{figure}

In this section, we extend the incremental algorithm to develop a batch algorithm that efficiently updates the truss decomposition after a batch of edge insertions. 
Our batch algorithm is motivated by the following observation.
Consider the graph in Figure \ref{fig4}, where the red edges form a 3-truss. 
Upon inserting the edge $e_1$, the \textit{HCQTY} version of the incremental algorithm explores all edges in red as part of \textsc{AlgorithmX}($3$), but none of the edges have their $K(.)$ values increase from 3 to 4. 
When we next add the edge $e_2$, the algorithm explores the same set of red edges along with the edge $e_1$. This time however, the red edges and the edges $e_1$ and $e_2$ form a 4-truss. As a result, the $K(.)$ values of all the edges are updated to 4. 
This example illustrates that the incremental algorithm upon inserting $e_1$ does no useful work, while the same algorithm upon inserting $e_2$ does the same work, but this time it does something useful. The batch algorithm we propose avoids performing these redundant computations.

%The central idea of the batch algorithm is a modification of \textsc{Property 1}, which we will state below as \textsc{Property 2}.\\
%\underline{\textsc{Property 2:}} The set of edges $S$ with $K(.)=k$ have their $K(.)$ values increase to $k+1$ \textit{if and only if} the edges in $S$ and a set of new edges $E$, together with a set $Q$ of edges with $K(.)>k$ form a subgraph where each edge supports at least $(k-1)$ triangles. \\

%As before, this property follows from the definition of a $k$-truss. Here, the set of new edges $E$ is a subset of the batch of edges $B$. Using this property, we develop the batch algorithm as follows.

Its central idea is \textsc{Property~1} mentioned in \textsc{Section~V}. 
Given a batch of edges $B$, the algorithm adds all the edges in $B$ to the graph and sets their initial $K(.)$ values to 2. 
Then it iteratively increases the $K(.)$ values of these edges till it computes their correct values, as follows.
First it picks an edge, say $e$, from this batch, and uses \textsc{Theorem 3} to explore the initial set of edges $S$, whose $K(.)$ values can increase from 2 to 3. Note that this set of edges can also include edges from $B$. 
As before, it then prunes the set $S$ to $S'$, before it increases the $K(.)$ values of all the edges in $S'$. As a result, every edge in the batch, as well as in the original graph that forms a 3-truss with the edge $e$ has its $K(.)$ value set to at least 3. 
The algorithm similarly checks every other edge in $B$, to see if can be part of a 3-truss. At the end of this iteration, every edge in $B$ that belongs to a 3-truss has its $K(.)$ increase from 2 to 3. 
In the next iteration, we similarly check for all the edges in $B$ with $K(.)=3$, if they could be part of a 4-truss. We continue in this fashion and appropriately increase the $K(.)$ of edges in $B$ from $k$ to $k+1$ for all $k \in [2,ktmax]$.

The batch algorithm leads to computational savings because every set of edges in $B$ that are part of the same $(k+1)$-truss have their $K(.)$ values increase from $k$ to $k+1$ at the same time. This is not the case when using the incremental algorithm that considers one edge at a time, as illustrated in the example above. Since both the incremental algorithm as well as the batch algorithm check for edges whose $K(.)$ values increase from $k$ to $k+1$ for all  $k \in [2,ktmax]$, the batch algorithm performs at most as many computations as the incremental algorithm.

%%%%%%%%%%%%%%%%%%%%%%%%%%%%%%%%%%%%%%%%%%%%%%%%%%%%%%%%%%%%%%%%%%%%%%%%%%%%%%%%%%%%%%%%%%

\begin{figure*}[t!]
    \centering
    \subfloat[\label{fig5:a}]{%
		\includegraphics[scale=0.42]{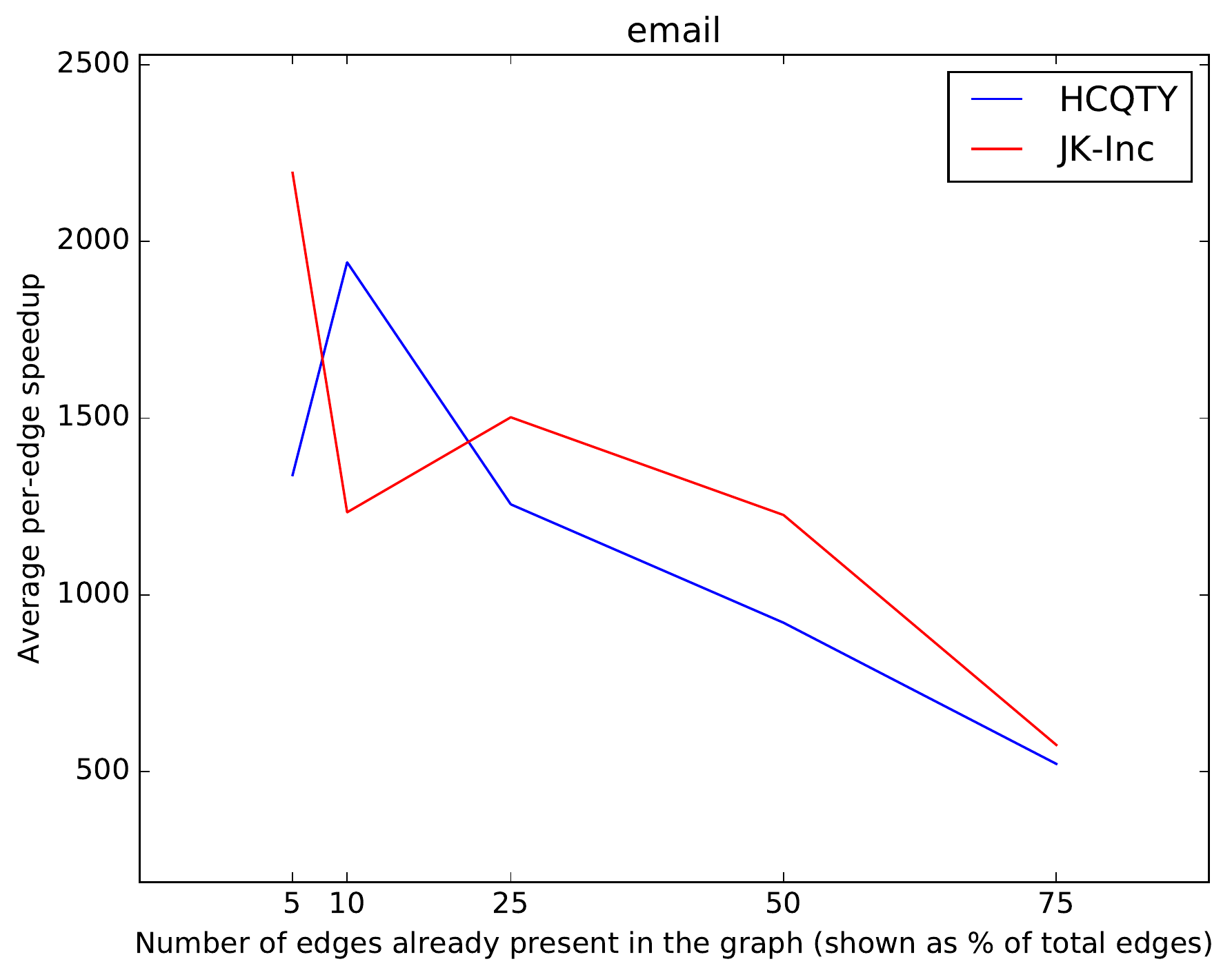}
    }
    \subfloat[\label{fig5:b}]{%
      	\includegraphics[scale=0.42]{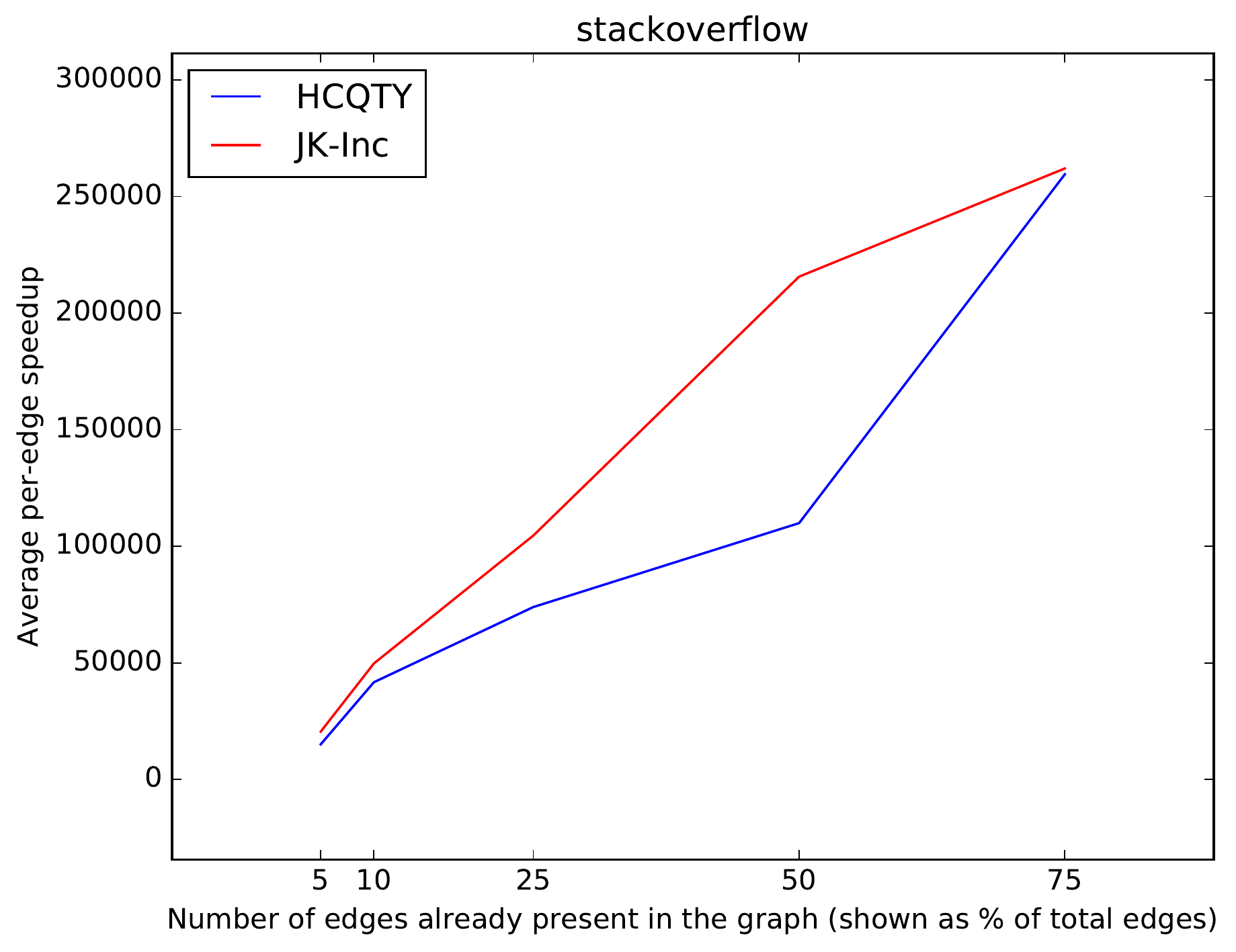}
    }
    \caption{The above plots show how the incremental algorithms scale for both dense (\texttt{email}) and sparse (\texttt{stackoverflow}) graphs. We construct static graphs with the first 5\%, 10\%, 25\%, 50\%, and 75\% of the temporal edges, for both the datasets. 
    The plot \ref{fig5:a} shows the average per-edge speedup over 100 edge insertions to the static graphs of \texttt{email}. 
    The plot \ref{fig5:b} show the same over 1000 edge insertions to the static graphs of \texttt{stackoverflow}.}
    \label{fig5}
\end{figure*}

\begin{figure*}[t!]
    \centering
    \subfloat[\label{fig6:a}]{%
		\includegraphics[scale=0.42]{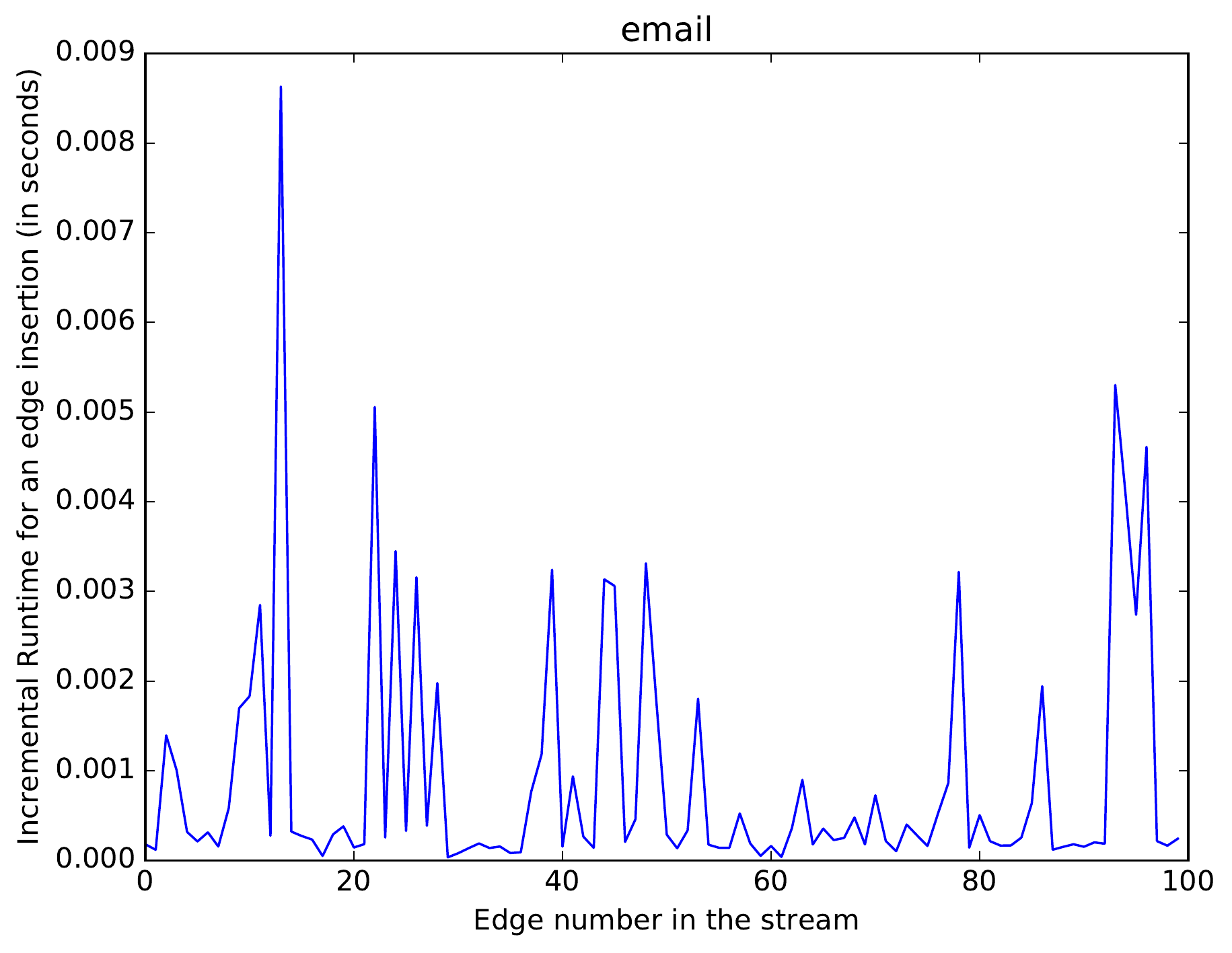}
    }
    \subfloat[\label{fig6:b}]{%
      	\includegraphics[scale=0.42]{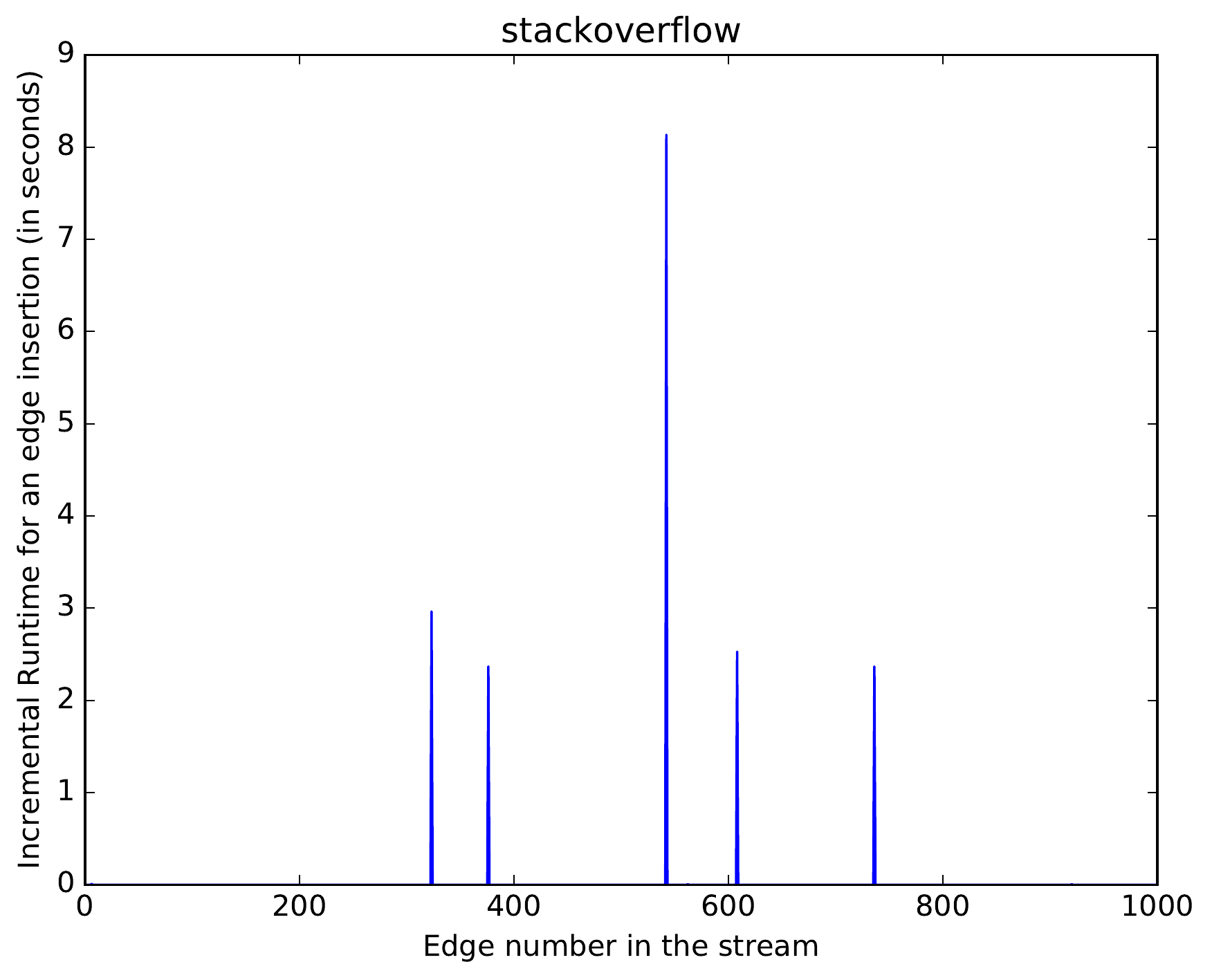}
    }
    \caption{In the above plots, we build the static graph with the first 75\% temporal edges, for both the datasets. 
    The plot \ref{fig6:a} shows the incremental runtime (of the \textsc{JK} version) in seconds when we add the next 100 temporal edges to the static graph of \texttt{email}. 
    Similarly, the plot \ref{fig6:b} shows the incremental runtime in seconds when we add the next 1000 temporal edges to the static graph of  \texttt{stackoverflow}.}
    \label{fig6}
\end{figure*}

\section{Experimental Methodology}
% We first perform a series of experiments to evaluate both the incremental algorithms $-$ the \textsc{HCQTY} version and the \textsc{JK} version $-$ among each other, and against the non-incremental algorithm. We then perform another set of experiments to evaluate the performance of the batch algorithm with respect to the incremental algorithms.

%\begin{figure*}
%    \centering
%    \subfloat[\label{fig5:a}]{%
%      \includesvg[scale=0.45]{images/small-per-edge-speedup-plot.svg}%
%    }
%    \subfloat[\label{fig5:b}]{%
%      \includesvg[scale=0.45]{images/large-per-edge-speedup-plot.svg}%
%    }
%    \caption{Plot (a) shows the average per-edge speedup over 100 edges, when added to the static graph built with the first 5\%, 10\%, 25\%, 50\% and 75\% of the edges of \texttt{email-Eu-core-temporal}. 
%    Plot (b) shows a similar plot for \texttt{sx-stackoverflow-a2q} with the average taken over 1000 edges. 
%    In both the plots, the \textit{blue} curve corresponds to \textit{v1} of the incremental algorithm, while \textit{red} curve corresponds to the \textit{v2} of the incremental algorithm.}
%    \label{fig5}
%\end{figure*}

\subsection{Datasets}
% We evaluate the performance of our algorithms on real-world temporal datasets available on SNAP, where each edge has a time-stamp associated with it to denote the time at which the edge was created. We present the experimental results for two datasets, which we denote with \texttt{stackoverflow} and \texttt{email}. 
% While \texttt{stackoverflow} refers to the \texttt{sx-stackoverflow-a2q} dataset on SNAP, which is a large and sparse network with 2464606 nodes and 17823525 temporal edges, \texttt{email} refers to the \texttt{email-Eu-core-temporal} dataset, which is a comparatively denser network with 986 nodes and 332334 temporal edges. 
We evaluated the performance of the algorithms on the \texttt{sx-stackoverflow-a2q} (\texttt{stackoverflow}) and \texttt{email-Eu-core-temporal} (\texttt{email}) datasets that are available in SNAP~\cite{leskovec2016snap}. 
Both of these datasets correspond to graphs whose edges have timestamps indicating when they appeared.
The \texttt{stackoverflow} dataset is a large sparse graph with 2464606 nodes and 17823525 temporal edges, and the \texttt{email} dataset is a comparatively denser graph with 986 nodes and 332334 temporal edges. 
These datasets have self-loops (an edge connecting a vertex to itself) and some edges could occur multiple times with different timestamps. We ignore such degenerate cases $-$ we ignore self-loops and consider an edge between 2 vertices only once. Moreover, we do not care about the edges being directed, and consider all edges to be undirected.

\subsection{Experimental Setup}
The experiments were conducted on a system with an eight-core \texttt{Intel Xeon E5-2640 v2} processor, 62GB of main memory and 20MB of last-level cache. 
Our algorithms are implemented in \texttt{C} and compiled with \texttt{gcc 5.4.0}. 

% In order to simulate streaming data, we first sort the edges according to their timestamp, and then build a static graph using the edges of the graph up to a selected timestamp. We then add edges after this timestamp one at a time to the graph.
We performed two sets of experiments. The first was to evaluate the performance of the incremental algorithm to update the truss decomposition after adding a single edge and the second was to evaluate the performance after adding a batch of edges.
In order to simulate streaming data, we first sorted the edges according to their timestamp, and then built a static graph using the edges of the graph up to a selected timestamp.
In the experiments to evaluate the performance of the incremental algorithm, we added edges after this timestamp one at a time, whereas in the experiments for batch algorithm, we added all the edges in the batch after the timestamp together.

In both sets of experiments, we evaluated the scalability of the algorithms as well. 
We built static graphs with the first 5\%, 10\%, 25\%, 50\% and 75\% of the edges sorted by timestamp, for both the datasets.
For the dense dataset, we then inserted the next 100 edges to each of the five static graphs. 
For the sparse dataset, we inserted the next 1000 edges. We increased the number of inserted edges for the sparse dataset for a more accurate analysis, since most edges in the sparse dataset do not affect the truss decomposition.

\subsection{Metrics}
We use the average per-edge speedup to evaluate the performance of the incremental algorithms (the \textit{HCQTY} version and the \textit{JK-Inc} version). 
When an edge is inserted, we take the ratio of the time taken to calculate the truss decomposition from scratch using the non-incremental algorithm and the time taken to update the truss decomposition using each of the incremental algorithms, to calculate their respective per-edge speedups.
We take the average of these speedups over a number of edges to calculate the average per-edge speedups.

To evaluate the performance of the batch algorithm with that of the incremental algorithm that adds one edge at a time, we compare their corresponding runtimes on adding a batch of edges.

\subsection{Methods compared}
We evaluated the performance of the following methods:
\begin{enumerate}
	\item \textit{Non-incremental}: In this algorithm, the truss decomposition is recomputed from scratch after every edge addition using the optimized serial peeling algorithm in~\cite{smith2017truss}. The efficiency of this algorithm is due to several optimizations with respect to triangle enumeration, which is a major cost during the peeling process.
	\item \textit{HCQTY}: This is our implementation of the algorithm presented in Huang et al.
	\item \textit{JK-Inc}: This is our incremental algorithm that is described in \textsc{Section V}.
	\item \textit{JK-Batch}: This is our batch algorithm that is described in \textsc{Section VI}.
	
\end{enumerate}
% \subsection{Setup for scalability experiments}
% In the first set of experiments, we compare the incremental algorithms based on their average per-edge speedup when compared to the non-incremental algorithm. 
% When an edge is inserted, we compare the time taken to update the truss decomposition using each of the incremental algorithms, versus the time taken to calculate the truss decomposition from scratch using the non-incremental algorithm, to calculate the per-edge speedup. 
% We take the average of these speedups over a number of edges to calculate the  average per-edge speedup in using these incremental algorithms.

% In order to perform scalability experiments, we build static graphs with the first 5\%, 10\%, 25\%, 50\% and 75\% of the edges sorted by timestamp, for both the datasets.
% For the dense dataset, we then insert the next 100 edges to each of the five static graphs. 
% For the sparse dataset, we insert the next 1000 edges. We increase the number of inserted edges for the sparse dataset for a more accurate analysis, since most edges in the sparse dataset do not affect the truss decomposition.
% We use this setup for evaluating both the incremental algorithms, where we insert the edges one at a time, as well as for evalulating the batch algorithm where we insert all the edges at once.

\section{Experimental Results}
\begin{figure*}[h]
    \centering
    \subfloat[\label{fig7:a}]{%
		\includegraphics[scale=0.42]{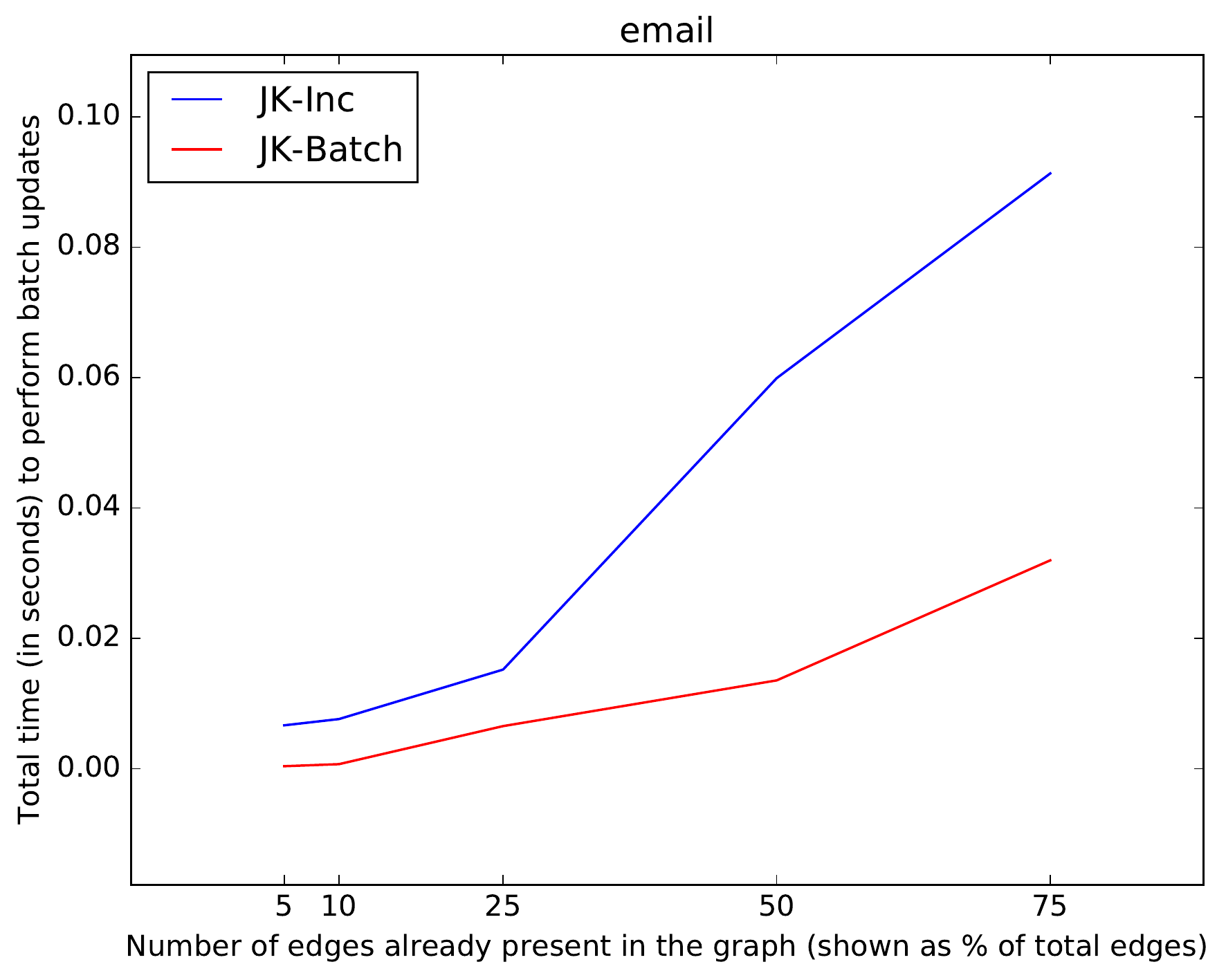}
    }
    \subfloat[\label{fig7:b}]{%
      	\includegraphics[scale=0.42]{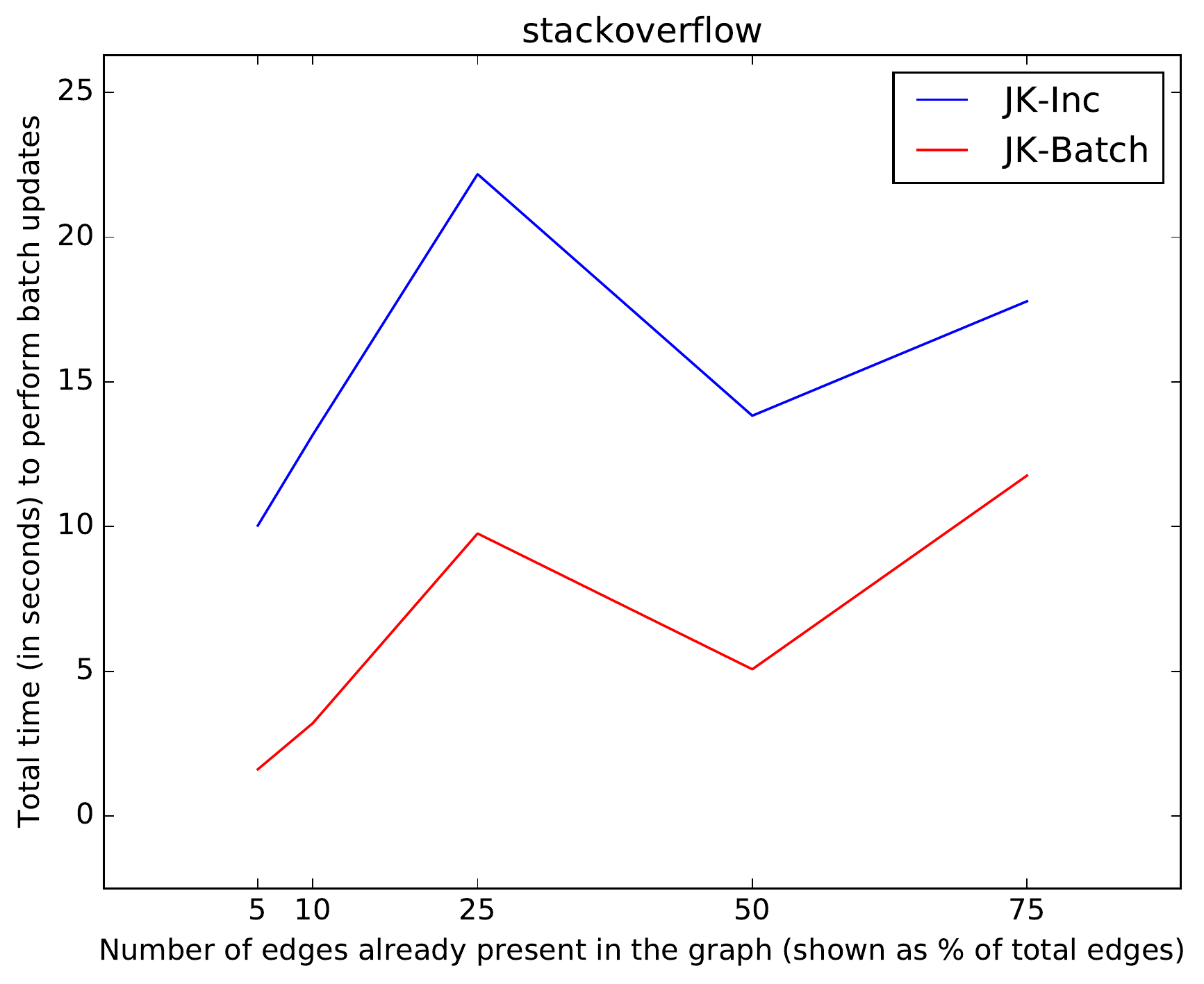}
    }
    \caption{The above plots show how the batch algorithm (\textit{JK-Batch}) performs as compared to the incremental algorithm (\textit{JK-Inc}), for both dense (\texttt{email}) and sparse (\texttt{stackoverflow}) graphs. 
    The plot \ref{fig7:a} shows the runtimes in seconds when we add 100 edges to the different static graphs of \texttt{email}. 
    The plot \ref{fig7:b} similarly shows the runtimes in seconds when we add 1000 edges to the different static graphs of \texttt{stackoverflow}. 
    }
    \label{fig7}
\end{figure*}

\subsection{Inter-evaluation of the incremental algorithms}
We compare \textit{JK-Inc} with \textit{HCQTY} based on their respective average per-edge speedups when compared to the \textit{non-incremental} algorithm. 
We also evaluate how the algorithms individually scale as the number of edges in the static graph increases.

The plots in Fig. \ref{fig5} show the performance of both incremental algorithms on our datasets. For the sparse dataset, \textit{JK-Inc} consistently performs better than \textit{HCQTY}.
This is because \textit{JK-Inc} explores a smaller set $S - $ the set of edges whose $K(.)$ values can potentially increase $-$ as compared to \textit{HCQTY}, and evicts a fewer set of edges to find the required set $S' - $ the set of edges whose $K(.)$ values actually increase $-$ thereby reducing the number of computations compared to \textit{HCQTY}.

For the dense dataset, the performance of \textit{JK-Inc} is again better than \textit{HCQTY}, except in one case. This is because in some cases the additional overhead of updating the memoized truss-degree of edges is expensive enough to worsen the performance of \textit{JK-Inc}. 
In general, if the set $S$ explored by \textit{JK-Inc} is not considerably smaller than the set $S$ explored by \textit{HCQTY}, then \textit{JK-Inc} performs more computations due to the overhead mentioned above.

\subsection{Performance of the incremental algorithms as the size of the graph increases}
Fig. \ref{fig5} shows that both \textit{JK-Inc} and \textit{HCQTY} scale similarly for each of the datasets as the size of the underlying graph increases. 
For the dense dataset, as the size of the static graph increases, the average per-edge speedup decreases. 
Since the graph is dense, any inserted edge has the potential to be part of multiple trusses that span a considerable portion of the graph. As a result, the set $S$ explored tends to be large. As more edges are added, the graph gets denser, thereby exacerbating the above effect, leading to a reduction in the average speedup.

In contrast, for the sparse dataset, the average  speedup increases as the size of the graph increases $-$ to as high as 250000 at 75\%. Since the dataset is a sparse graph, any inserted edge is likely to be part of only a few trusses, most of which span only a small portion of the graph. This trend does not change much as we increase the size of the static graph, since the entire graph itself is sparse. 
As the size of the graph increases, the non-incremental algorithm does increasingly more work, whereas the incremental algorithm explores a smaller fraction of the entire graph.
As a result, the performance of the incremental algorithm is particularly suitable for large, sparse graphs, which is the common characteristic of most real-world datasets.

\subsection{Analysis of per-edge incremental update time}
In this section, we assess the time associated with each incremental update to the truss decomposition as different edges are inserted. 
Since the previous section suggests that ~\textit{JK-Inc} performs better than \textit{HCQTY} in most cases, we perform the analysis for only \textit{JK-Inc} $-$ all conclusions drawn are valid for \textit{HCQTY} as well.
We look at the runtimes of ~\textit{JK-Inc} as we add edges to a static graph built with the first 75\% of the edges. We do this for both the datasets to analyze the performance of \textit{JK-Inc} in sparse as well as in dense graphs.

For the dense graph, the plot in Fig. \ref{fig6:a} shows a plot with lots of spikes, which suggests that incremental algorithm explores a substantial number of edges for most edge insertions, while for others there is negligible amount of work done. 
The reason for this follows from the previous discussion $-$ any inserted edge has a considerable chance of affecting a large portion of the graph. 

In contrast, the plot corresponding to the sparse graph in Fig. \ref{fig6:b} has only five spikes over 1000 edge insertions. 
Again, it follows from the previous discussion that any inserted edge in the sparse graph is unlikely to affect a large portion of the graph; for those inserted edges that do affect a large portion of the graph, the incremental algorithm performs a lot of computations. 
The incremental algorithms involves more computations per edge as compared to the non-incremental algorithms.
As a result, if the incremental algorithm explores a large enough portion of the graph such that the incremental algorithm performs more computations than the non-incremental algorithm, it would be better to use the non-incremental algorithm to update the truss decomposition.

\subsection{Performance of the batch algorithm}
We compare the performance of \textit{JK-Batch} with the performance of \textit{JK-Inc}.
The plots in Figure \ref{fig7} show that the batch algorithm always performs better than the incremental algorithm.
This is as expected, since the batch algorithm performs at most as many computations as the incremental algorithm. 
When a set of edges in the batch are part of the same truss, the batch algorithm updates the truss numbers of all the edges belonging to that truss at the same time. In contrast, the incremental algorithm performs the same amount of computation once for each edge in the set. 

For the \texttt{email} dataset with 5\% of the edges in the static graph, inserting a batch of the next 100 edges using \textit{JK-Inc} takes 0.00667 seconds while using \textit{JK-Batch} takes 0.00038 seconds, providing a speedup of upto 17.5. For the \texttt{stackoverflow} dataset, \textit{JK-Batch} provides a speedup of upto 6 as compared to \textit{JK-Inc} when inserting a batch of the next 1000 edges.

%In the next set of experiments, we further analyze our incremental algorithms by analyzing the runtime for each edge in a batch of edges. 
%This gives a deeper understanding of the algorithms and provides insights into any existing bottlenecks. 
%We do this by building the static graph for both datasets, consisting of the first 75\% of the edges sorted by the timestamp, and adding 100 edges to the dense dataset, and 1000 edges to the sparse dataset. 

%Finally, we use the setup for scalability experiments to evaluate the performance of our batch algorithm.

%%%%%%%%%%%%%%%%%%%%%%%%%%%%%%%%%%%%%%%%%%%%%%%%%%%%%%%%%%%%%%%%%%%%%%%%%%%%%%%%%%%%%%%%%%%%%%%%
\section{Conclusion}
In this paper, we first developed a theory that identifies a set of edges whose truss numbers can potentially change upon an edge insertion. Based on this theory, we then develop an algorithm similar to the one proposed by Huang et al.\cite{huang2014querying}, which we call as the \textit{HCQTY} version. 
We then improved this algorithm by incorporating certain optimizations. We call this version as the \textit{JK-Inc} version.

Further, we showed that some parts of the algorithm are independent of each other, that can be exploited for parallelism. However, we have not provided implementation details and experimental analysis for this.

Then, we extended the theory behind the incremental algorithms to perform batch updates, and developed the first batch algorithm for truss decomposition.

We then performed a series of experiments to compare the two incremental algorithms, and found that the \textit{JK-Inc} version performs better than the \textit{HCQTY} version in general. 
We further show that the incremental algorithms scale well for sparse graphs, but not as well for dense graphs. Since most real-world graphs tend to be large and sparse in nature, using the incremental algorithms in such cases is beneficial.
Our experiments on batch updates show that the batch algorithm always performs better than the incremental algorithm.

In addition, as evidenced by the experiments performed, the incremental algorithms take a considerable amount of time in some cases. In situations like this, we might want to revert to using the non-incremental algorithm $-$ such an approach requires having to predict beforehand if the incremental algorithm would perform worse that the non-incremental algorithm. 
	Moreover, if the batch size is large enough, the batch algorithm would perform worse than using the non-incremental algorithm to recompute the truss decomposition from scratch after all the edges in the batch are inserted. We do not perform experiments to analyze this behavior and obtain the optimal batch sizes in different scenarios.
We wish to explore the above mentioned ideas as part of our future work. \\

\bibliographystyle{IEEEtran}
\bibliography{IEEEabrv,references.bib}


% \end{thebibliography}

\end{document}